\documentclass[conference]{IEEEtran}

\usepackage{amsmath,graphicx}
\usepackage{amssymb}
\usepackage{amsfonts}
\usepackage{amsthm}
\usepackage{mathrsfs}
\usepackage{subfig}

\newtheorem{theorem}{Theorem}

\theoremstyle{remark}

\theoremstyle{definition}

\ifCLASSINFOpdf
\else
\fi
\begin{document}
\title{Parameter Tracking via Optimal Distributed Beamforming \\in 
an Analog Sensor Network}
%\author{\IEEEauthorblockN{Michael Shell}
%\IEEEauthorblockA{School of Electrical and\\Computer Engineering\\
%Georgia Institute of Technology\\
%Atlanta, Georgia 30332--0250\\
%Email: http://www.michaelshell.org/contact.html}
%\and
%\IEEEauthorblockN{Homer Simpson}
%\IEEEauthorblockA{Twentieth Century Fox\\
%\and
%\IEEEauthorblockN{James Kirk\\ and Montgomery Scott}
%\IEEEauthorblockA{Starfleet Academy\\
%San Francisco, California 96678-2391\\


\author{\IEEEauthorblockN{
Feng Jiang, Jie Chen 
and
A. Lee Swindlehurst}
\IEEEauthorblockA{
Center for Pervasive Communications and Computing (CPCC)\\
Department of EECS, University of California at Irvine\\
Irvine, CA 92697, USA\\
\{feng.jiang, jie.chen, swindle\}@uci.edu}
}


\maketitle

\begin{abstract}
We consider the problem of optimal distributed beamforming in a sensor
network where the sensors observe a dynamic parameter in noise and
coherently amplify and forward their observations to a fusion center
(FC).  The FC uses a Kalman filter to track the parameter using the
observations from the sensors, and we show how to find the optimal
gain and phase of the sensor transmissions under both global and
individual power constraints in order to minimize the mean squared
error (MSE) of the parameter estimate.  For the case of a global power
constraint, a closed-form solution can be obtained.  A numerical
optimization is required for individual power constraints, but the
problem can be relaxed to a semidefinite programming problem (SDP),
and we show how the optimal solution can be constructed from the
solution to the SDP.  Simulation results show that compared with equal
power transmission, the use of optimized power control can
significantly reduce the MSE.
\end{abstract}


% \fi
\IEEEpeerreviewmaketitle

\section{Introduction}\label{sec:intro}
In an analog-based distributed sensor network, the sensor nodes
multiply their noisy observations by a gain and phase and transmit the
result to a fusion center (FC).  The FC then uses the sum of the
received signals to estimate the parameter.  The key problem in this
setting is to design the optimal gain and phase multiplier for each
sensor in order to obtain the most accurate parameter estimate at the
FC.  Furthermore, these multipliers must be updated in situations
where the parameter is time-varying.  Some examples of prior work on
this type of problem include
\cite{Cui:2007,Xiao:2008,jiang:2011,Leong:2011,Leong:20112}.  In
\cite{Cui:2007}, an orthogonal multiple access channel (MAC) was
assumed between the sensor nodes and FC.  The FC used a best linear
unbiased estimator to estimate a static parameter and and the optimal
power allocation with both sum and individual power constraints were
investigated to minimize the mean square error (MSE). A coherent MAC
was considered in \cite{Xiao:2008} and a linear minimum mean square
error estimator was adopted at the FC to estimate the Gaussian
source. The optimal power allocation problem was solved under a total
transmit power constraint. A phase-only optimization problem was
formulated in \cite{jiang:2011} and the phase of the transmitted
signal from different sensor nodes was adjusted such that the received
signal at the FC can be added coherently to optimize the performance
of a maximum likelihood (ML) estimator. In \cite{Leong:2011} and
\cite{Leong:20112}, the parameter of interest was modeled as a dynamic
process and the FC employed a Kalman filter to track the parameter. In \cite{Leong:2011}, a
power optimization problem was formulated to minimize the MSE under a
sum power constraint and an asymptotic expression for the outage
probability of the MSE was derived for a large number of sensor nodes.
Additionally, the problem of minimizing MSE outage probability was studied in \cite{Leong:20112}.

In this paper, we consider a setup similar to \cite{Leong:2011} and
\cite{Leong:20112}. We assume that the parameter of interest is a
dynamic process and the sensor nodes coherently amplify (gain and
phase) and forward their observations of the parameter to the FC.  The
sensors act like a distributed beamformer, but they are also
forwarding their observation of the background noise along with the
measured parameter.  The FC uses a Kalman filter to track the dynamic
process, and we show how the transmit gain and phase of the sensor can
be optimized at each time step to minimize the MSE of the parameter
estimate.  We assume that the optimized gain and phase is fed back to
the sensor from the FC at each step, prior to the next measurement.
The contributions of this paper are as follows:
\begin{enumerate}
\item We find a closed-form solution for the optimal transmit gain and
phase that minimizes MSE under a sum power constraint. This problem
was also solved in \cite{Leong:2011} using the KKT conditions derived
in \cite{Xiao:2008}.  However, our approach converts the problem to a
Rayleigh quotient maximization problem and results in a simpler and
more direct solution.
\item The problem of minimizing the MSE under individual sensor power
constraints is solved by relaxing it to a semi-definite programming
(SDP) problem, and then proving that the optimal solution can be
constructed from the SDP solution.
\item For a suboptimal case where the sensor nodes use equal power
transmission, we derive an exact expression for the MSE outage probability.
%of MSE, and show that an increase in sum transmit power results in an
%exponential decrease in the outage probability.
\end{enumerate}

\section{System Model}\label{sec:model}
We model the complex-valued dynamic parameter $\theta_n$ as a
first-order Gauss-Markov process:
\begin{equation}
\theta_n=\alpha\theta_{n-1}+u_n\nonumber\;,
\end{equation}
where the process noise $u_n$ has distribution
$\mathcal{CN}(0,\sigma_u^2)$. Assuming the FC and the sensor node are
all configured with a single antenna, the received signal at the FC is
\begin{equation}\label{eq:observ}
y_n=\mathbf{a}_n^H\mathbf{h}_n\theta_n+\mathbf{a}_n^H\mathbf{H}_n\mathbf{v}_n+w_n\;,
\end{equation}
where $\mathbf{h}_n=[h_{1,n},\dots,h_{N,n}]^T$ and
$h_{i,n}\in\mathbb{C}$ is the channel coefficient between the $i$th
sensor and the FC, $\mathbf{a}_n=[a_{1,n},\dots,a_{N,n}]^{T}$ is the
conjugate of the sensor transmit gain and phase,
$\mathbf{H}_n=\mathrm{diag}\{h_{1,n},\dots,h_{N,n}\}$, $\mathbf{v}_n$
is Gaussian measurement noise at the sensors with covariance
$\mathbf{V}=\mathbb{E}\{\mathbf{v}_n\mathbf{v}_n^{H}\}=\mathrm{diag}\left\{\sigma_{v,1}^2,\cdots,\sigma_{v,N}^2\right\}$,
and $w_n$ is additive white Gaussian noise at the fusion center with
variance $\sigma_{w}^2$. The channel parameter is defined as
\begin{equation}
h_{i,n}=\frac{\tilde{h}_{i,n}}{d_{i}^\gamma}\;\nonumber,
\end{equation}
where $\tilde{h}_{i,n}$ is complex Gaussian with zero
mean and unit variance, $d_i$ denotes the distance between sensor $i$
and the FC, and $\gamma$ is the path-loss exponent.

Based on the above dynamic and observation models, the standard
Kalman Filter is defined by the following quantities:
\begin{itemize}
\item Prediction Step: $\hat{\theta}_{n|n-1}=\alpha\hat{\theta}_{n-1|n-1}$

\item Prediction MSE:
$P_{n|n-1}=\alpha^2P_{n-1|n-1}+\sigma_u^2$
\item Kalman Gain:
\begin{equation}
k_n=\frac{P_{n|n-1}\mathbf{h}_n^H\mathbf{a}_n}{\mathbf{a}_n^H\mathbf{H}_n\mathbf{V}\mathbf{H}_n^H\mathbf{a}_n+P_{n|n-1}\mathbf{a}_n^H\mathbf{h}_n\mathbf{h}_n^H\mathbf{a}_n+\sigma_w^2}\nonumber
\end{equation}
\item Measurement Update:
\begin{equation}
\hat{\theta}_{n|n}=\hat{\theta}_{n|n-1}+k_n\left(y_n-\mathbf{a}_n^H\mathbf{h}_n\hat{\theta}_{n|n-1}\right)\nonumber
\end{equation}
\item MSE:
\begin{equation}\label{eq:mse}
P_{n|n}=(1-k_n\mathbf{a}_n^H\mathbf{h}_n)P_{n|n-1}\;.
\end{equation}
\end{itemize}
\section{Minimizing MSE under Global Power Constraint}\label{sec:mse1}
In this section, we formulate and solve the problem under the
assumption that the sensor nodes have a sum power constraint.
The optimization problem is formulated as
\begin{eqnarray}\label{eq:opt1}
\min_{\mathbf{a_n}} && P_{n|n}\\
s. t. &&\mathbf{a}_n^H\mathbf{D}\mathbf{a}_n\le P_{\max}\nonumber\;,
\end{eqnarray}
where
$\mathbf{D}=\textrm{diag}\{\sigma_{\theta}^2+\sigma_{v,1}^2,\cdots,
\sigma_{\theta}^2+\sigma_{v,N}^2 \}$, $\sigma_{\theta}^2$ denotes the variance of $\theta_n$ and $P_{\max}$ is the maximum
sum transmit power.  From (\ref{eq:mse}), to minimize MSE $P_{n|n}$,
we need to maximize $k_n\mathbf{a}_n^H\mathbf{h}_n$ which is
calculated as
\begin{eqnarray}
&&k_n\mathbf{a}_n^H\mathbf{h}_n=\frac{P_{n|n-1}\mathbf{a}_n^H\mathbf{h}_n\mathbf{h}_n^H\mathbf{a}_n^H}{\mathbf{a}_n\mathbf{H}_n\mathbf{V}\mathbf{H}^H_n\mathbf{a}_n+P_{n|n-1}\mathbf{a}_n^H\mathbf{h}_n\mathbf{h}_n^H\mathbf{a}_n+\sigma_w^2}\;.\nonumber
\end{eqnarray}
Thus, the optimization problem (\ref{eq:opt1}) is equivalent to 
\begin{eqnarray}\label{eq:opt2}
\max_{\mathbf{a}_n} &&\frac{\mathbf{a}_n^H\mathbf{h}_n\mathbf{h}_n^H\mathbf{a}_n}{\mathbf{a}_n^H\mathbf{H}_n\mathbf{V}\mathbf{H}^H_n\mathbf{a}_n+\sigma_w^2}\\
s. t. &&\mathbf{a}_n^H\mathbf{D}\mathbf{a}_n\le P_{\max}\;.\nonumber
\end{eqnarray}

Denote the optimal solution to problem (\ref{eq:opt2}) as
$\mathbf{a}^{*}_n$. It is easy to verify that the sum transmit
power constraint should be met with equality
$\mathbf{a}^{*H}_n\mathbf{D}\mathbf{a}^{*}_n=P_{\max}$, so that~(\ref{eq:opt2}) 
can be rewritten as
\begin{eqnarray}\label{eq:opt3}
\max_{\mathbf{a}_n} &&\frac{\mathbf{a}_n^H\mathbf{h}_n\mathbf{h}_n^H\mathbf{a}_n}{\mathbf{a}_n^H(\mathbf{H}_n\mathbf{V}\mathbf{H}^H_n+\frac{\sigma_w^2}{P_{\max}}\mathbf{D})\mathbf{a}_n}\\
s. t. &&\mathbf{a}_n^H\mathbf{D}\mathbf{a}_n=P_{\max}\; . \nonumber
\end{eqnarray}
Problem (\ref{eq:opt3}) maximizes a Rayleigh quotient under
a quadratic constraint, which results in a simple closed-form
solution. If we define
$\mathbf{B}=\mathbf{H}_n\mathbf{V}\mathbf{H}^H_n+\frac{\sigma_w^2}{P_{\max}}\mathbf{D}$,
the optimal solution is given by
\begin{equation}\label{eq:optimal}
\mathbf{a}_n^{*}=\sqrt{\frac{P_{\max}}{\mathbf{h}_n^H\mathbf{B}^{-1}\mathbf{D}^{-1}\mathbf{B}^{-1}\mathbf{h}_n}}\mathbf{B}^{-1}\mathbf{h}_n\nonumber
\end{equation}
and the optimal value of (\ref{eq:opt3}) is calculated as  
\begin{eqnarray}\label{eq:maxvalue}
\frac{\mathbf{a}_n^{*H}\mathbf{h}_n\mathbf{h}_n^H\mathbf{a}_n^{*}}{\mathbf{a}_n^{*H}(\mathbf{H}_n\mathbf{V}\mathbf{H}^H_n+\frac{\sigma_w^2}{P_{\max}}\mathbf{D})\mathbf{a}_n^{*}}&=&\mathbf{h}_{n}^{H}\mathbf{B}^{-1}\mathbf{h}_n \; ,\nonumber
\end{eqnarray}
which is a random variable that depends on the distribution of
$\mathbf{h}_n$.  An upper bound for
$\mathbf{h}_{n}^{H}\mathbf{B}^{-1}\mathbf{h}_n$ is given by
\begin{eqnarray}
\mathbf{h}_{n}^{H}\mathbf{B}^{-1}\mathbf{h}_n&\overset{(a)}{<}&\mathbf{h}_{n}^{H}(\mathbf{H}\mathbf{V}\mathbf{H}^H)^{-1}\mathbf{h}_n\nonumber\\
&=&\sum_{i=1}^N\frac{1}{\sigma_{v,i}^2}\;,
\end{eqnarray}
where $(a)$ follows from $\mathbf{B}^{-1}\prec(\mathbf{H}\mathbf{V}\mathbf{H}^H)^{-1}$. Plugging (\ref{eq:maxvalue}) into (\ref{eq:mse}), we obtain a lower bound on the MSE:
\begin{eqnarray}\label{eq:lb}
P_{n|n}&>&\left(1-\frac{1}{1+\frac{1}{\left(\sum_{i=1}^{N}\frac{1}{\sigma_{v,i}^2}\right)P_{n|n-1}}}\right)P_{n|n-1}\nonumber\\
&=&\frac{P_{n|n-1}}{1+\left(\sum_{i=1}^{N}\frac{1}{\sigma_{v,i}^2}\right)P_{n|n-1}}\;.\nonumber
\end{eqnarray}
This lower bound can be asymptoticly achieved with $P_{\max}\to\infty$ or $\sigma_w^2\ll \sigma_{v,i}^2$, and the corresponding sensor transmit gain and phase is
\begin{equation}\label{eq:optimal2}
\mathbf{a}_{n}^{*}\!=\!\!\sqrt{\frac{P_{\max}}{\sum_{i=1}^N\frac{1}{\sigma_{v,i}^2(\sigma_{\theta}^2+\sigma_{v,i}^2)}}}\left[\frac{1}{\bar{h}_{1,n}\sigma_{v,1}^2},\cdots, \frac{1}{\bar{h}_{N,n}\sigma_{v,N}^2}\right].
\end{equation}
From (\ref{eq:optimal2}), it can be observed that sensors whose
product $|h_{i,n}|\sigma_{v,i}^2$ is small will be allocated more
transmit power.

\section{Minimizing MSE under Individual Power Constraints}\label{sec:mse2}

When the sensor nodes have individual power constraints, the optimal
distributed beamforming problem becomes
\begin{eqnarray}\label{eq:opt4}
\max_{\mathbf{a}_n} &&\frac{\mathbf{a}_n^H\mathbf{h}_n\mathbf{h}_n^H\mathbf{a}_n}{\mathbf{a}_n^H\mathbf{H}_n\mathbf{V}\mathbf{H}^H_n\mathbf{a}_n+\sigma_w^2}\\
s. t. && |a_{i,n}|^2(\sigma_{\theta}^2+\sigma_{v,i}^2)\le P_{\max,i}\;,\quad i=1,\cdots, N\nonumber\;,
\end{eqnarray}
where $P_{\max,i}$ is the maximum transmit power of the $i$th sensor
node. Problem (\ref{eq:opt4}) is a quadratically constrained ratio of
two quadratic functions (QCRQ).  Using the approach proposed in
\cite{Beck:2010}, the QCRQ problem can be relaxed to an SDP
problem. Introduce a real auxiliary variable $t$ and define
$\tilde{\mathbf{a}}_{n}=t\mathbf{a}_n$, so that problem (\ref{eq:opt4})
is equivalent to
\begin{eqnarray}\label{eq:opt5}
\max_{\mathbf{a}_n, t} &&\frac{\tilde{\mathbf{a}}_n^H\mathbf{h}_n\mathbf{h}_n^H\tilde{\mathbf{a}}_n}{\tilde{\mathbf{a}}_n^H\mathbf{H}_n\mathbf{V}\mathbf{H}^H_n\tilde{\mathbf{a}}_n+\sigma_w^2t^2}\\
s. t. &&\tilde{\mathbf{a}}_n^H\mathbf{D}_i\tilde{\mathbf{a}}_n\le t^2P_{\max,i}\;,\quad i=1,\cdots, N\;,\nonumber\\
      &&t\neq 0\;,\nonumber
\end{eqnarray}
where $\mathbf{D}_i=\textrm{diag}\{0,\cdots,0,
\sigma_{\theta}^2+\sigma_{v,i}^2,0,\cdots,0\}$.  Then, we can further
rewrite problem (\ref{eq:opt5}) as
\begin{eqnarray}\label{eq:opt6}
\max_{\mathbf{a}_n, t} &&\tilde{\mathbf{a}}_n^H\mathbf{h}_n\mathbf{h}_n^H\tilde{\mathbf{a}}_n\\
s. t. &&\tilde{\mathbf{a}}_n^H\mathbf{H}_n\mathbf{V}\mathbf{H}^H_n\tilde{\mathbf{a}}_n+\sigma_w^2t^2=1\;,\nonumber\\
      &&\tilde{\mathbf{a}}_n^H\mathbf{D}_i\tilde{\mathbf{a}}_n\le t^2P_{\max,i},\quad i=1,\cdots, N\;.\nonumber
\end{eqnarray}
Note that the constraints in problem (\ref{eq:opt6}) already guarantee
that $t\neq 0$, so the constraint $t\neq 0$ is removed.

Define $\bar{\mathbf{a}}_n=[\tilde{\mathbf{a}}^{H}_n, t]^{H}$, $\bar{\mathbf{H}}_n=\left[\begin{array}{cc}\mathbf{h}_n\mathbf{h}_n^H&0\\
                \mathbf{0}^T&0
                \end{array}\right]$, $\bar{\mathbf{C}}_n=\left[\begin{array}{cc}\mathbf{H}_n\mathbf{V}\mathbf{H}^H_n&0\\
                \mathbf{0}^T&\sigma_w^2
                \end{array}\right]$, and $\bar{\mathbf{D}}_i=\left[\begin{array}{cc}\mathbf{D}_i&0\\
                \mathbf{0}^T&-P_{\max,i}
               \end{array}\right]$, so that problem (\ref{eq:opt6}) can be written in the compact form
\begin{eqnarray}\label{eq:opt7}
\max_{\bar{\mathbf{a}}_n} &&\bar{\mathbf{a}}_n^H\bar{\mathbf{H}}_n\bar{\mathbf{a}}_n\\
s. t. &&\bar{\mathbf{a}}_n^H\bar{\mathbf{C}}_n\bar{\mathbf{a}}_n=1\;,\nonumber\\
      &&\bar{\mathbf{a}}_n^H\bar{\mathbf{D}}_i\bar{\mathbf{a}}_n\le 0\;,\quad i=1,\cdots, N\;. \nonumber
\end{eqnarray}
Problem (\ref{eq:opt7}) is equivalent to 
\begin{eqnarray}\label{eq:opt8}
\max_{\bar{\mathbf{A}}}&&\mathrm{tr}(\bar{\mathbf{A}}\bar{\mathbf{H}}_n)\\
s. t. &&\mathrm{tr}(\bar{\mathbf{A}}\bar{\mathbf{C}}_n)=1\;, \nonumber\\
&&\mathrm{tr}(\bar{\mathbf{A}}\bar{\mathbf{D}}_i)\le 0\;,\quad i=1,\cdots, N\;,\nonumber\\
&&\mathrm{rank}(\bar{\mathbf{A}})=1\;.\nonumber
\end{eqnarray}

By relaxing the rank-one constraint on $\bar{\mathbf{A}}$, we convert
problem (\ref{eq:opt8}) to a standard SDP problem:
\begin{eqnarray}\label{eq:opt9}
\max_{\bar{\mathbf{A}}}&&\mathrm{tr}(\bar{\mathbf{A}}\bar{\mathbf{H}}_n)\\
s. t. &&\mathrm{tr}(\bar{\mathbf{A}}\bar{\mathbf{C}}_n)=1\;, \nonumber\\
&&\mathrm{tr}(\bar{\mathbf{A}}\bar{\mathbf{D}}_i)\le 0\;,\quad i=1,\cdots, N\;,\nonumber\\
&&\bar{\mathbf{A}}\succeq 0\;.\nonumber
\end{eqnarray}
The above problem can be solved in polynomial time using the interior
point method. Due to the relaxation of the rank-constraint on
$\bar{\mathbf{A}}$, the optimal value of problem (\ref{eq:opt9})
provides an upper bound for problem (\ref{eq:opt4}). After obtaining
the optimal solution $\bar{\mathbf{A}}^{*}$, a rank-one solution
$\mathbf{a}^{*}_n$ can be recovered for the original problem
(\ref{eq:opt4}). In the following, we show that based on
$\bar{\mathbf{A}}^{*}$ a rank-one solution $\mathbf{a}^{*}_n$ can be
constructed such that $\mathbf{a}^{*}_n$ is the optimal solution to
problem (\ref{eq:opt4}).

Defining $\bar{\mathbf{A}}^{*}_{l,m}$ as the $(l,m)$th element of
$\bar{\mathbf{A}}^{*}$ and $\bar{\mathbf{A}}_N^{*}$ as the $N$th order
leading principal submatrix of $\bar{\mathbf{A}}^{*}$ formed by
deleting the $(N+1)$st row and column of $\bar{\mathbf{A}}^{*}$, we
propose the following theorem:

\begin{theorem}\label{theorem1}
Define the optimal solution to problem (\ref{eq:opt9}) as
$\bar{\mathbf{A}}^{*}$, then
$\bar{\mathbf{A}}^{*}_N=\mathbf{a}\mathbf{a}^{H}$ and the optimal
solution to problem (\ref{eq:opt4}) is given by
$\mathbf{a}^{*}_n=\frac{1}{\sqrt{\bar{\mathbf{A}}_{N+1,N+1}^{*}}}\mathbf{a}$\;.
\end{theorem}
\begin{IEEEproof}
We first utilize the strong duality between problem (\ref{eq:opt9})
and its dual problem to show the property of the optimal solution
$\bar{\mathbf{A}}^{*}$. The dual problem of problem (\ref{eq:opt9}) is
given by \cite{Zhang:2011}:
\begin{eqnarray}\label{eq:opt10}
\min_{y_i,z}&& z\\
s. t. &&\sum_{i=1}^{N}y_i\bar{\mathbf{D}}_i+z\bar{\mathbf{C}}_n-\bar{\mathbf{H}}_n\succeq 0\;, \nonumber\\
&&y_1,\dots,y_N,z\ge 0\nonumber\;.
\end{eqnarray}
It is easy to verify that there exists strictly feasible points for
problem (\ref{eq:opt9}) and problem (\ref{eq:opt10}). For problem
(\ref{eq:opt9}), we can construct
\begin{equation}
\bar{\mathbf{A}}^f=\textrm{diag}\{ab, \cdots, ab, b\}\;,\nonumber
\end{equation}
where
$0<a<\min_{i}\frac{P_{\max,i}}{\sigma_{\theta}^2+\sigma_{v,i}^2}$ and
$b=\frac{1}{\sum_{i=1}^{N}a|h_{n,i}|^2\sigma_{v,i}^2+\sigma_w^2}$.  For
problem (\ref{eq:opt10}), we can randomly select $y_i^f>0$, and set
$z^f$ large enough such that
\begin{equation} z^f\!>\max\left\{\!\frac{\mathbf{h}_{n}^{H}\mathbf{h}_n\!+\!\!\sum_{i=1}^Ny_i^fP_{\max,i}}{\sigma_w^2},\frac{\mathbf{h}_n^{H}\mathbf{h}_n\!-\!y_i^f(\sigma_{\theta}^2\!+\!\sigma_{v,i}^2)}{|h_{n,i}|^2\sigma_{v,i}^2}\!\right\}\;.\nonumber
\end{equation}
Then, according to Slater's theorem, strong duality holds between the primal problem (\ref{eq:opt9}) and the dual problem (\ref{eq:opt10}) and we have the following complementary condition:
\begin{equation}\label{eq:slater}
\textrm{tr}(\bar{\mathbf{A}}^{*}\mathbf{G}^{*})=0\;,
\end{equation}
where $\mathbf{G}^{*}=\sum_{i=1}^{N}y_i^{*}\bar{\mathbf{D}}_i+z^{*}\bar{\mathbf{C}}_n-\bar{\mathbf{H}}_n$ and $y_{i}^{*}$ and $z^{*}$ denote the optimal solution to problem (\ref{eq:opt10}). 
Due to the special structure of $\bar{\mathbf{D}}_i$, $\bar{\mathbf{C}}_n$ and $\bar{\mathbf{H}}_n$, $\mathbf{G}^{*}$ can be expressed as
\begin{equation}
\mathbf{G}^{*}=\left[\begin{array}{cc}\mathbf{G}^{*}_N&0\\
                \mathbf{0}^T&\mathbf{G}^{*}_{N+1,N+1}
                \end{array}\right]\;,\nonumber
\end{equation}
where $\mathbf{G}^{*}_N=\sum_{i=1}^Ny_i^*\mathbf{D}_i+z^*\mathbf{H}_n\mathbf{V}\mathbf{H}_n^H-\mathbf{h}_n\mathbf{h}_n^H$ and $\mathbf{G}^{*}_{N+1,N+1}=z^{*}\sigma_w^2-\sum_{i=1}^{N}y_i^{*}P_{\max,i}$.
Since both $\bar{\mathbf{A}}^{*}$ and $\mathbf{G}^{*}$ are positive semidefinite, Eq. (\ref{eq:slater}) is equivalent to 
\begin{equation}
\bar{\mathbf{A}}^{*}\mathbf{G}^{*}=0\;.\nonumber
\end{equation}
\iffalse
\begin{equation}
\mathbf{G}^{*}_{N+1,N+1}=0
\end{equation}
\fi
Additionally, with consideration of the structure of $\mathbf{G}^{*}$, we have
\begin{equation}
\bar{\mathbf{A}}^{*}_N\mathbf{G}^{*}_N=0\;.\nonumber
\end{equation}

Define $\mathbf{V}_{G}$ as a set of vectors
orthogonal to the row space of 
$\mathbf{G}^{*}_{N}$.  Then the column vectors of
$\bar{\mathbf{A}}_{N}^{*}$ must belong to span($\mathbf{V}_G$) and
$\textrm{rank}(\bar{\mathbf{A}}^{*}_N)\le
\textrm{rank}(\mathbf{V}_G)$.  Given two matrices $\mathbf{M}$ and
$\mathbf{N}$, we have
$\textrm{rank}(\mathbf{M}+\mathbf{N})\ge|\textrm{rank}(\mathbf{M})-\textrm{rank}(\mathbf{N})|$
\cite{Gentle:2007}, thus, a lower bound for
$\textrm{rank}(\mathbf{G}_N^{*})$ is calculated as
\begin{eqnarray}
\textrm{rank}(\mathbf{G}_N^{*})\!\!\!&\ge&\!\!\!\textrm{rank}\left(\sum_{i=1}^Ny_i^*\mathbf{D}_i+z^*\mathbf{H}_n\mathbf{V}\mathbf{H}_n^H\right)\!\!-\textrm{rank}(\mathbf{h}_n\mathbf{h}_n^H)\nonumber\\
&=&\!\!\!N-1\;.\nonumber
\end{eqnarray}
An upper bound for $\textrm{rank}(\mathbf{V}_G)$ is then given by
\begin{eqnarray}\label{eq:rank1}
\textrm{rank}(\mathbf{V}_G)&=&N-\textrm{rank}(\mathbf{G}_N^{*})\\
&\le&1\nonumber\;.
\end{eqnarray}
Since $\text{tr}(\bar{\mathbf{A}}^{*}\bar{\mathbf{H}})=\mathbf{h}_n^H\bar{\mathbf{A}}^*_N\mathbf{h}_n$ and $\text{tr}(\bar{\mathbf{A}}^{*}\bar{\mathbf{H}})>\text{tr}(\bar{\mathbf{A}}^{f}\bar{\mathbf{H}})>0$, we have 
\begin{equation}\label{eq:rank2}
\bar{\mathbf{A}}^*_N\neq 0 \qquad \textrm{rank}(\bar{\mathbf{A}}^*_N)\ge 1\;.
\end{equation} 
Combining Eqs. (\ref{eq:rank1}) and (\ref{eq:rank2}), we have
\begin{equation}
\textrm{rank}(\bar{\mathbf{A}}^*_N)=1\;.\nonumber
\end{equation}

Define the rank-one decomposition of $\bar{\mathbf{A}}_N^{*}$ as
$\bar{\mathbf{A}}_N^{*}=\mathbf{a}\mathbf{a}^{H}$, so that the optimal
rank-one solution to problem (\ref{eq:opt9}) is
\begin{equation}
\bar{\mathbf{a}}^{*}=[\mathbf{a}^{H}, \sqrt{\bar{\mathbf{A}}_{N+1,N+1}^*}]^{H}\;.\nonumber
\end{equation}
If the optimal solution of problems (\ref{eq:opt4}) is $\mathbf{a}^{*}_n$, 
then 
\begin{equation}
\frac{\mathbf{a}_n^{*H}\mathbf{h}_n\mathbf{h}_n^H\mathbf{a}_n^{*}}{\mathbf{a}_n^{*H}\mathbf{H}\mathbf{V}\mathbf{H}^H\mathbf{a}_n^{*}+\sigma_w^2}\le\textrm{tr}(\bar{\mathbf{A}}^{*}\bar{\mathbf{H}})\;,\nonumber
\end{equation} 
where equality can be achieved provided that an optimal rank-one
solution exists for problem (\ref{eq:opt9}). Since
$\textrm{tr}(\bar{\mathbf{A}}^{*}\bar{\mathbf{D}}_i)\le 0$,
$\bar{\mathbf{A}}^{*}\neq 0$ and $\mathbf{D}_i\succ 0$, then we have
$\bar{\mathbf{A}}^{*}_{N+1,N+1}>0$, otherwise
$\textrm{tr}(\bar{\mathbf{A}}^{*}\bar{\mathbf{D}}_i)> 0$, which
contradicts the constraints in problem (\ref{eq:opt9}).  Based on
$\bar{\mathbf{a}}^{*}$, the optimal solution to problem
(\ref{eq:opt4}) is given by
\begin{equation}
\mathbf{a}^{*}_n=\frac{1}{\sqrt{\bar{\mathbf{A}}_{N+1,N+1}^*}}\mathbf{a}\;,\nonumber
\end{equation}
and we have
\begin{equation}
\frac{\mathbf{a}_n^{*H}\mathbf{h}_n\mathbf{h}_n^H\mathbf{a}_n^{*}}{\mathbf{a}_n^{*H}\mathbf{H}\mathbf{V}\mathbf{H}^H\mathbf{a}_n^{*}+\sigma_w^2}=\textrm{tr}(\bar{\mathbf{A}}^{*}\bar{\mathbf{H}})\;,\nonumber
\end{equation}
which verifies the optimality of $\mathbf{a}^{*}_n$.
\end{IEEEproof}
\iffalse
In addition to the optimal numerical solution provided in Theorem \ref{theorem1}, we attempt to calculate approximate closed-form solution. When $P_{\max,i}\gg \sigma_w^2$, at the FC, the additive noise's effect can be neglected and the value of the objective function only depends on the direction of the power control parameter $\mathbf{a}$, and the optimal direction of $\mathbf{a}$ is given by Eq. (\ref{eq:lb}). To satisfy the individual power constraint, we need to scale $\mathbf{a}^{*}$ properly, and the approximate solution is 
\begin{equation}
\mathbf{a}_{n}^{*}\!=\!\!p\left[\frac{1}{\bar{h}_{1,n}\sigma_{v,1}^2},\cdots, \frac{1}{\bar{h}_{N,n}\sigma_{v,N}^2}\right],
\end{equation}
where $p=\min_i|h_{i,n}|\sigma_{v,i}^2\sqrt{\frac{P_{\max,i}}{\sigma_{v,i}^2+\sigma_{\theta}^2}}$.\fi

\section{Equal Power Allocation}\label{sec:conpower}
Here we calculate the MSE outage probability of a suboptimal
solution in which each sensor transmits with the same power.
The outage probability derived here can serve as an upper
bound for the outage performance of the optimal algorithm
with individual power constraints. The transmit gain is given by
\begin{equation}
\mathbf{a}_{e}=\sqrt{\frac{P_{\max}}{N}}\left[\frac{1}{\sqrt{\sigma_{\theta}^2+\sigma_{v,1}^2}},\cdots, \frac{1}{\sqrt{(\sigma_{\theta}^2+\sigma_{v,N}^2)}}\right]\;.\nonumber
\end{equation}
For this suboptimal approach, the MSE is calculated as
{\small \begin{equation}
P_{n|n}=\left(\!1-\frac{P_{n|n-1}\mathbf{a}_e^H\mathbf{h}_n\mathbf{h}_n^H\mathbf{a}_e}{\mathbf{a}_e^H\mathbf{H}_n\mathbf{V}\mathbf{H}_n\mathbf{a}_e+P_{n|n-1}\mathbf{a}_e^H\mathbf{h}_n\mathbf{h}_n^H\mathbf{a}_e+\sigma_w^2}\right)P_{n|n-1}\nonumber\;,
\end{equation}}which is a random variable depending on the distribution of the channel parameter $\mathbf{h}_n$. Define the outage probability as $P_{out}=\mathrm{Pr}\left\{P_{n|n}>\epsilon\right\}$, so that
\begin{eqnarray}
P_{out}&=&\mathrm{Pr}\left\{\frac{\mathbf{a}_e^H\mathbf{h}_n\mathbf{h}_n^H\mathbf{a}_e}{\mathbf{a}_e^H\mathbf{H}_n\mathbf{V}\mathbf{H}^H_n\mathbf{a}_e+\sigma_{w}^2}<\beta\right\}\nonumber\\
&=&\mathrm{Pr}\left\{\mathbf{a}_e^H\mathbf{h}_n\mathbf{h}_n^H\mathbf{a}_e-\beta\mathbf{a}_e^H\mathbf{H}_n\mathbf{V}\mathbf{H}^H_n\mathbf{a}_e<\beta\sigma_w^2\right\}\nonumber\\
&=&\mathrm{Pr}\left\{\tilde{\mathbf{h}}_n^H\left(\bar{\mathbf{D}}\mathbf{a}_e\mathbf{a}_e^H\bar{\mathbf{D}}-\beta\mathbf{E}\right)\tilde{\mathbf{h}}_n\le \beta\sigma_w^2\right\},\nonumber
\end{eqnarray}
where $\beta=\frac{P_{n|n-1}-\epsilon}{\epsilon P_{n|n-1}}$,
$\bar{\mathbf{D}}=\mathrm{diag}\left\{\frac{1}{d_{1}^{\gamma}},\cdots,
\frac{1}{d_{N}^{\gamma}}\right\}$,
$\mathbf{E}=\mathrm{diag}\left\{\frac{P_{\max}\sigma_{v,i}^2}{N(\sigma_{\theta}^2+\sigma_{v,i}^2)d_{i}^{2\gamma}},\cdots,\frac{P_{\max}\sigma_{v,N}^2}{N(\sigma_{\theta}^2+\sigma_{v,N}^2)d_{i}^{2\gamma}}\right\}$,
$\tilde{\mathbf{h}}_n=[\tilde{h}_{1,n},\cdots,
\tilde{h}_{N,n}]$. 

Define
$\mathbf{B}=\bar{\mathbf{D}}\mathbf{a}_e\mathbf{a}_e^H\bar{\mathbf{D}}-\beta\mathbf{E}$,
and denote the eigenvalues of $\mathbf{B}$ as
$\lambda_1,\cdots,\lambda_{N}$, then the random variable
$\tilde{\mathbf{h}}_{n}^H\mathbf{B}\mathbf{h}_{n}$ can be viewed as
the weighted sum of independent chi-square random variables
$\sum_{i}^{N}\lambda_i\chi_i(2)$. Based on the results in
\cite{Al_Naffouri:2009}, we have
\begin{equation}\label{eq:cdf}
P_{out}=1-\sum_{i=1}^{N}\frac{\lambda_i^N}{\prod_{l\neq i}(\lambda_i-\lambda_l)}\frac{1}{|\lambda_i|}e^{-\frac{(P_{n|n-1}-\epsilon)\sigma_w^2}{\epsilon P_{n|n-1}\lambda_i}}u(\lambda_i)\;,
\end{equation}
where $u(\cdot)$ is the unit step function.  If we let $e_{1}\ge\cdots
\ge e_{N}$ denote the eigenvalues of $\mathbf{E}$, then from Weyl's
inequality \cite{So:1999} we have the following bounds:
\begin{eqnarray}
\mathbf{a}_{e}^H\bar{\mathbf{D}}^2\mathbf{a}_e-\beta e_{1}\le\lambda_{1}&\!\!\!\!\le\!\!\!\!\!&\mathbf{a}_{e}^H\bar{\mathbf{D}}^2\mathbf{a}_e-\beta e_{N}\label{eq:bound}\nonumber\\
-\beta e_{N-i+1}\le\lambda_{i}&\!\!\!\!\!\le\!\!\!\!&-\beta e_{N-i+2}\;,\quad 2\le i\le N\;, \label{eq:bound2}\nonumber
\end{eqnarray}
where
$\mathbf{a}_{e}^H\bar{\mathbf{D}}^2\mathbf{a}_{e}=\sum_{i=1}^{N}\frac{P_{\max}}{N(\sigma_{\theta}^2+\sigma_{v,i}^2)d_{i}^{2\gamma}}$. Since
only $\lambda_1$ can be positive, equation~(\ref{eq:cdf}) can be simplified as
\begin{equation}
P_{out}=\left\{\begin{array}{lr}
1-\frac{\lambda_1^{N-1}}{\prod_{l\neq 1}(\lambda_1-\lambda_l)}e^{-\frac{(P_{n|n-1}-\epsilon)\sigma_w^2}{\epsilon P_{n|n-1}\lambda_1}}\;,&\textrm{$\lambda_1>0$}\;,\\
1\;,&\textrm{$\lambda_1\le 0$}\;.\nonumber
\end{array} \right.
\end{equation}
%Since $\lambda_i\propto P_{\max}$, we conclude that increasing the sum
%transmit power can exponentially reduce the outage probability
%$P_{out}$. 
Since it is not possible to evaluate the $\lambda_i$ in
closed-form, the above outage probability expression must be calculated
numerically.

\section{Simulation Results}\label{sec:simu}
\begin{figure}
\centering
\includegraphics[height=3in, width=3.8in]{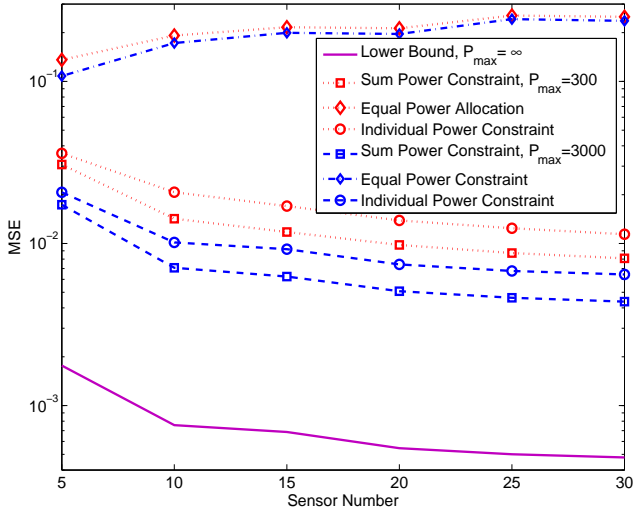}
\caption{MSE vs. number of sensors for $P_{\max}=300$ or $3000$.}\label{f1}
\end{figure}

\begin{figure}
\centering
\includegraphics[height=3in, width=3.8in]{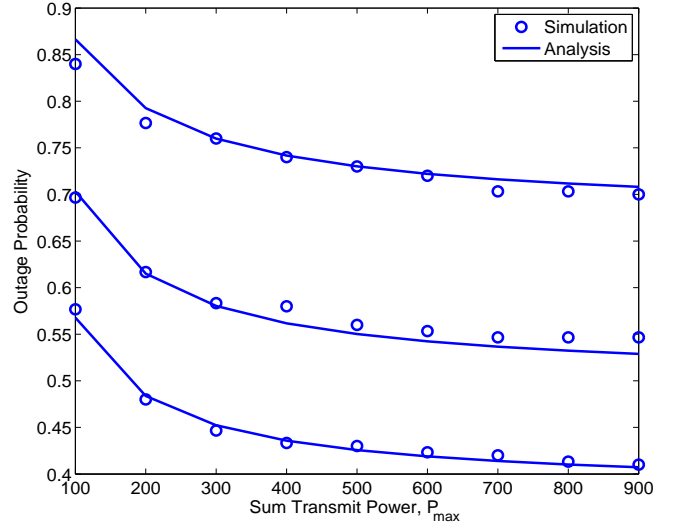}
\caption{MSE outage probability for equal power allocation vs.
sum transmit power for $N=10$ sensors.}\label{f2}
\end{figure}

To verify the performance of the proposed optimization approaches, the
results of several simulation examples are described here. In the
simulation, the distance to the sensors $d_i$ is uniformly distributed
over $[2,8]$ and the path loss exponent $\gamma$ is set to 1. The variance $\sigma_{\theta}^2$ is set to 1, and the $P_{n|n-1}$ is initialized as 1.
The MSE is obtained by averaging over 300 realizations of $\mathbf{h}_n$. The
observation noise power $\sigma_{v,i}^2$ is uniformly distributed over
$[0, 0.5]$ and the power of the additive noise at the FC is set to
$\sigma_w^2=0.5$. Two different sum power constraints are considered
$P_{\max}=300$ or $3000$. To fairly compare the results under the sum
power constraint and the individual power constraint, we set
$P_{\max,i}=\frac{P_{\max}}{N}$. In Fig.~\ref{f1}, the results show
that compared with equal power allocation, the optimized power
allocation significantly reduces the MSE.  In fact, adding sensors
with equal power allocation actually increases the MSE, while the MSE
always decreases for the optimal methods.  The extra flexibility of
the global power constraint leads to better performance compared with
individual power constraints, but the difference is not large.  The
lower bound shows the performance that could be achieved with
$P_{\max}\to\infty$.  The theoretical and simulated outage
probabilities of the equal power allocation is presented in Fig
{\ref{f2}}. The results show that the theoretical analysis matches well
with the simulations.  

\section{Conclusion}\label{sec:conclusion}\label{sec:conc}
In this paper, we considered optimal distributed beamforming for an
analog sensor network attempting to track a dynamic parameter under
both global and individual power constraints.  For the sum power
constraint case, we derived a closed-form solution for the optimal
sensor transmit gain and phase.  For individual power constraints, we
developed a numerical optimization procedure that is guaranteed to
find the optimal sensor gains and phases.  We also derived an exact
expression for the MSE outage probability of a suboptimal scheme in
which each sensor transmits with equal power.  Simulations were
presented to verify the performance of the optimal algorithms and the
accuracy of the MSE outage probability expression.

\bibliography{reference}

\begin{thebibliography}{10}
\providecommand{\url}[1]{#1}
\csname url@samestyle\endcsname
\providecommand{\newblock}{\relax}
\providecommand{\bibinfo}[2]{#2}
\providecommand{\BIBentrySTDinterwordspacing}{\spaceskip=0pt\relax}
\providecommand{\BIBentryALTinterwordstretchfactor}{4}
\providecommand{\BIBentryALTinterwordspacing}{\spaceskip=\fontdimen2\font plus
\BIBentryALTinterwordstretchfactor\fontdimen3\font minus
  \fontdimen4\font\relax}
\providecommand{\BIBforeignlanguage}[2]{{%
\expandafter\ifx\csname l@#1\endcsname\relax
\typeout{** WARNING: IEEEtran.bst: No hyphenation pattern has been}%
\typeout{** loaded for the language `#1'. Using the pattern for}%
\typeout{** the default language instead.}%
\else
\language=\csname l@#1\endcsname
\fi
#2}}
\providecommand{\BIBdecl}{\relax}
\BIBdecl

\bibitem{Cui:2007}
S.~Cui, J.-J. Xiao, A.~J. Goldsmith, Z.-Q. Luo, and H.~V. Poor, ``Estimation
  diversity and energy efficiency in distributed sensing,'' \emph{IEEE Trans.
  Signal Process.}, vol.~55, no.~9, pp. 4683--4695, Sep. 2007.

\bibitem{Xiao:2008}
J.-J. Xiao, S.~Cui, Z.-Q. Luo, and A.~J. Goldsmith, ``Linear coherent
  decentralized estimation,'' \emph{IEEE Trans. Signal Process.}, vol.~56,
  no.~2, pp. 757--770, Feb. 2008.

\bibitem{jiang:2011}
F.~Jiang, J.~Chen, and A.~L. Swindlehurst, ``Phase-only analog encoding for a
  multi-antenna fusion center,'' in \emph{Proc. IEEE ICASSP 2012}, March 2012,
  pp. 2645--2648.

\bibitem{Leong:2011}
A.~S. Leong, S.~Dey, G.~N. Nair, and P.~Sharma, ``Asymptotics and power
  allocation for state estimation over fading channels,'' \emph{IEEE Trans.
  Aero. and Elec. Sys.}, vol.~47, no.~1, pp. 611--633, Jan. 2011.

\bibitem{Leong:20112}
------, ``Power allocation for outage minimization in state estimation over
  fading channels,'' \emph{IEEE Trans. Signal Process.}, vol.~59, no.~7, pp.
  3382--3397, July 2011.

\bibitem{Beck:2010}
A.~Beck and M.~Teboulle, ``On minimizing quadratically constrained ratio of two
  quadratic functions,'' \emph{Journal of Convex Analysis}, vol.~17, no. 3 and
  4, pp. 789--804, 2010.

\bibitem{Zhang:2011}
Y.~Zhang and A.~M.-C. So, ``Optimal spectrum sharing in \protect{MIMO}
  cognitive radio networks via semidefinite programming,'' \emph{IEEE J. Sel.
  Areas Commun.}, vol.~29, no.~2, pp. 362--373, Feb. 2011.

\bibitem{Gentle:2007}
J.~E. Gentle, \emph{Matrix Algebra: Theory, Computations, and Applications in
  Statistics}.\hskip 1em plus 0.5em minus 0.4em\relax New York: Springer, 2007.

\bibitem{Al_Naffouri:2009}
T.~Y. Al-Naffouri and B.~Hassibi, ``On the distribution of indefinite quadratic
  forms in \protect{Gaussian} random variables,'' in \emph{Proc. IEEE ISIT
  2009}, Jun. 2009, pp. 1744--1748.

\bibitem{So:1999}
W.~So, ``Rank one perturbation and its application to the laplacian spectrum of
  a graph,'' \emph{Linear and Multilinear Algebra}, vol.~46, no.~3, pp.
  193--198, 1999.

\end{thebibliography}
\end{document}